\documentclass[12pt]{article}

\usepackage[body={17.5cm, 21cm},right=2cm]{geometry}

\usepackage{cool}
\usepackage{color}

\usepackage{graphicx}
\usepackage{epsf}
\usepackage{graphicx,epsfig}
\usepackage{amsmath} 
\usepackage{amsthm} 
\usepackage{amssymb}	
\usepackage{graphicx} 
\usepackage{multicol} 

\usepackage{amsmath}
\usepackage{amssymb}
\usepackage{epsfig}
\usepackage{cite}
\usepackage{color,colordvi}
\newcommand{\avg}[1]{\left< #1 \right>} 
\newcommand{\quadra}[1]{\left[ #1 \right]} 
\newcommand{\graffa}[1]{\left\{ #1 \right\}} 
\renewcommand{\d}[2]{\frac{d #1}{d #2}} 
 
\let\baraccent=\= 
\renewcommand{\=}[1]{\stackrel{#1}{=}} 

\newcommand{\be}{\begin{eqnarray}}
\newcommand{\ee}{\end{eqnarray}}
\newcommand{\bi}{\begin{itemize}}
\newcommand{\ei}{\end{itemize}}

\newcounter{hran}

\def\MSbar{\relax\ifmmode\overline{\rm MS}\else{$\overline{\rm MS}${ }}\fi}

\def\d{\rm d}

\def\d{{\rm d}}

\numberwithin{equation}{section}
\begin{document}\thispagestyle{empty}
\vspace{5mm}
\vspace{0.5cm}
\begin{center}

\def\thefootnote{\fnsymbol{footnote}}

{\Large \bf 
Renormalization of Composite Operators \\
\vspace{0.25cm}	
in time-dependent Backgrounds}
\\[2.5cm]
{\large  Simone Dresti    and Antonio Riotto}
\\[0.5cm]

\vspace{.3cm}
{\normalsize { \it Department of Theoretical Physics and Center for Astroparticle Physics (CAP)\\ 24 quai E. Ansermet, CH-1211 Geneva 4, Switzerland}}\\

\vspace{.3cm}

\vspace{.3cm}


\end{center}

\vspace{2cm}

\hrule \vspace{0.5cm}
{\small  \noindent \textbf{Abstract} \\[0.3cm]
\noindent 
We study  the phenomenon of composite operator renormalization and mixing in systems where
time-translational invariance is broken and the evolution is out-of-equilibrium. We  
show that composite operators mix also through  non-local memory terms which persist for periods whose duration is set by the 
mass scales in the problem.

\vspace{0.5cm}  \hrule
\vskip 1cm

\def\thefootnote{\arabic{footnote}}
\setcounter{footnote}{0}


\baselineskip= 15pt

\newpage 

\section{Introduction}\pagenumbering{arabic}
Out-of-equilibrium phenomena play a crucial  role during the evolution of the universe. They happen  soon after the  inflationary epoch \cite{lrreview},  during the reheating stage when the vacuum energy is converted into thermal particles, at the creation of the baryon asymmetry \cite{baureview,leptogenesis}, during  the freeze-out of dark matter particles \cite{reviewdm}, during  the formation  of light element abundances called  nucleosynthesis,  and at the generation of the cosmic microwave background radiation from the last scattering surface \cite{dodelson}.
In all these phenomena  one is interested in time-dependent
settings and  the objects to compute are  the time evolution
of the expectation values of observables rather than calculating scattering processes
using the $S$-matrix. The appropriate
formalism  is the so-called in-in formalism and was first developed by
Schwinger and Keldysh \cite{in1,in2,in3}. It allows to choose an arbitrary initial state and to follow its causal evolution consistently
including quantum effects. 

On the other hand, one is typically interested in observables, like the particle number and the
energy density of the system, which in quantum field theory are given by the expectation value of composite operators, that is of products of quantum fields evaluated at the same space point.  Due to the local product of quantum fields making up
the composite operator, new ultraviolet divergences appear. These divergences are generally not canceled
by the Lagrangian counter-terms and their  renormalization
 requires the introduction of new
counter-terms. An illustrative example of these new divergences is given in \cite{Ilderton} where the renormalization of composite operators is needed to extract the radiation reaction effects from QED. 
 Renormalized operators can be defined, which
are generically expressed  as linear combinations of all the
bare operators of equal or lower canonical dimensionality \cite{op1,op2}. In other words, composite operators mix with each other. This is the reason why, for instance, the definition of particle number density in an interacting
theory is a delicate matter. The necessity of giving a meaning
to divergent composite operators calls into play operator
mixing, so that a separation between different particle species
turns out, in general, to be a renormalization scale-dependent
procedure.

The scope of this paper is to explore the phenomenon of composite operators mixing in time-dependent set-ups. We will consider a simple Lagrangian made of two interacting scalar fields $\phi$ and $\chi$ and study the renormalization of the composite operators
$\phi^2$, $\chi^2$, and $\phi\chi$. We will show  that the mixing of the composite operators occur in a way different from what happens
in systems which are time-translation invariant. Indeed, the out-of-equilibrium evolution causes the appearance of   non-local (in time) kernels, thus 
introducing   memory effects in the system. These memory effects are indeed typical in quantum systems \cite{dan} and play a role in electroweak baryogenesis \cite{out1} and leptogenesis \cite{out2}. Indeed, the same non-local kernel appears
in the construction
of effective field theories for time-dependent systems evolving out-of-equilibrium \cite{holman}. Such new  terms cannot arise from a local action
of an effective field theory in terms of the light field, though they disappear in the adiabatic
limit. After a brief introduction to the in-in formalism in section 2, we will perform our calculations in two relatively simple time-dependent set-ups. The first, described in section 3, is in Minkowski space-time  where the time-translation breaking is introduced by a finite initial time $t_{\rm in}$. 
The second is the subject of section 4 and deals with a period of de Sitter to mimic what happens during the primordial stage of inflation. In both cases we find that memory effects appear in the mixing of the composite operators. These memories persist for a period whose duration 
is dictated by the mass scales involved. Having memory effects and different mass scales in the problem makes the process of diagonalization of composite operators more difficult than it is in time-translational set-ups. Our conclusions are contained in section 5.

\noindent
\section{The in-in formalism}\label{cap:SK}
In a Lorentz-invariant quantum field theory it is possible to define systems that break  time-translational invariance, for example through an explicit time-dependent Hamiltonian $H(t)$. In such a case  we are more interested in calculating expectation values of operators, rather than the traditional $S$-matrix elements. This motivates the use of so-called in-in formalism, which we  will briefly summarize in the following.

We consider a quantum system governed by a time-dependent Hamiltonian $H(t)$ in a state described by the density matrix $\rho(t)$. The expectation value of an observable $\langle\mathcal{O}(t)\rangle$ is given by
\begin{equation}\label{eq:<O(t)>}
\langle\mathcal{O}(t)\rangle = \text{Tr}\left[\rho(t)\mathcal{O}(t)\right].
\end{equation}
The time evolution of expectation values can be easily expressed in the interaction picture separating the  free and the interacting parts of the Hamiltonian, i.e. $H(t) = H_0(t) + H_I(t)$.
The density matrix evolves according to the Liouville equation

\begin{equation}\label{eq:Liouville}
\begin{cases}
i\dot{\rho}(t) = \left[H_I(t), \rho(t)\right], \\
\rho(t_\text{in}) = \rho_\text{in}
\end{cases}
\end{equation}
and can be solved by introducing the time-evolution operator $U_I(t, t_\text{in})$ as the solution of the Dyson equation

\begin{equation}\label{eq:Dyson}
\begin{cases}
i\dot{ U}_I(t,t_\text{in}) = H_I(t)U_I(t, t_\text{in}), \\
U_I(t_\text{in}, t_\text{in}) = \mathbb{I}.
\end{cases}
\end{equation}
Consequently $\rho(t)$ can be expressed in terms of $U$, $U^\dagger$ and the initial condition $\rho_\text{in}$ as

\begin{equation}\label{eq:rho(t)}
\rho(t)=U_I(t,t_\text{in})\rho_\text{in}U_I^\dagger(t,t_\text{in}) .
\end{equation}\\
To compute $\rho(t)$ it is therefore sufficient to find the solution of the Dyson equation (\ref{eq:Dyson}) which reads
\begin{equation}
U_I(t,t_\text{in})={\rm T} e^{-i\int_{t_\text{in}}^t {\rm d}\tau H_I(\tau)},
\end{equation}
where T stands for the time-ordered product (and ${\rm \overline{T}}$  to the anti-time-ordered product). The expression for $\rho$ follows immediately as

\begin{equation}
\rho(t)={{\rm T} e^{-i\int_{t_\text{in}}^t {\rm d}\tau H_I(\tau)}\,}\rho_\text{in}\, {{\rm \overline{T}} e^{+i\int_{t_\text{in}}^t {\rm d}\tau H_I(\tau)}} .
\end{equation}
\\
\\
This gives an explicit expression

\begin{equation}
\langle\mathcal{O}(t)\rangle =\text{Tr}\left\{\rho_\text{in} {{\rm \overline{T}} e^{+i\int_{t_\text{in}}^t {\rm d}\tau H_I(\tau)}} \mathcal{O}(t){{\rm T} e^{-i\int_{t_\text{in}}^t {\rm d}\tau H_I(\tau)}}\right\}.
\end{equation} 
The expression under the trace, reading from right to left, describes the evolution from the initial time $t_\text{in}$, where the initial density matrix is given, up to time $t$, where the observable $\mathcal{O}$ should be evaluated. Then one returns back to $t_\text{in}$. 
It is convenient to extend the time evolution to $t=+\infty$. A common trick is to insert $\mathbb{I}=U^\dagger_I(\infty,t)U_I(\infty,t)$ to the left of $\mathcal{O}(t)$, so that
\be
\avg{\mathcal{O}(t)} = \text{Tr}\graffa{\rho_\text{in} {{\rm \overline{T}}  e^{+i\int_{t_\text{in}}^{\infty} {\rm d}\tau H_I(\tau)}}{{\rm T}  e^{-i\int_{t}^\infty {\rm d}\tau H_I(\tau)}}    \mathcal{O}(t){{\rm T}  e^{-i\int_{t_\text{in}}^t {\rm d}\tau H_I(\tau)}}} .
\ee
This represents the time evolution along the closed time contour $\mathcal{C}$ shown in Fig. $\ref{fig:contour}$. We notice that the observable $\mathcal{O}(t)$ is evaluated in the forward part of the contour $\mathcal{C}$ because we have inserted the identity $U^\dagger_I(+\infty,t)U_I(+\infty,t)$ to the left of $\mathcal{O}(t)$. It is clear that one could have  inserted the identity to the right of $\mathcal{O}(t)$. In this case the same result is obtained with the exception that $\mathcal{O}(t)$ is evaluated in the backward part of the contour. 
\begin{figure}[htbp]
\centering
\includegraphics[width=0.6\columnwidth]{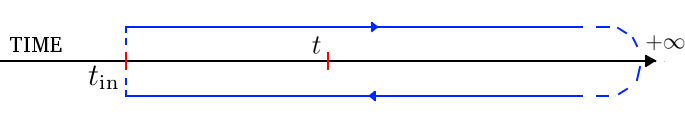}
\caption{\emph{{\small Closed time contour $\mathcal{C}$}}. \label{fig:contour}}
\end{figure}
Equivalently, one could also say that in the forward and backward parts of the contour $\mathcal{C}$, two different fields, let us call them generically $\phi_{\pm}$, propagate. Let us use the $+$ label for fields that propagate along  the forward part and are governed by $H^+(t)=H[\phi_+(\mathbf{x},t)]$ and $-$ for fields along the backward part governed by $H^-(t)=H[\phi_-(\mathbf{x},t)]$. Thus, $+$ fields evolve according to $U(+\infty, t_\text{in})$ and $-$ fields according to $U^\dagger(+\infty, t_\text{in})$. 
The expectation value (\ref{eq:<O(t)>}) can be expressed using the contour time-ordered product ${\rm T}_\mathcal{C}$
\be
\avg{\mathcal{O}(t)} = \text{Tr}\graffa{\rho_\text{in}\, {\rm T}_\mathcal{C} \,{\mathcal{O}^+(t)\,e^{-i\int_{t_\text{in}}^{+\infty} {\rm d} \tau \quadra{H_I^+(\tau)-H_I^-(\tau)}}}}.
\ee
Here  ${\rm T}_\mathcal{C}$ means that $+$ fields occur before $-$ fields and in the opposite order. 
It is worth mentioning at this moment that by choosing flat space standard modes as initial condition at finite $t_{\rm in}$, we are  making an assumption about the initial vacuum which is not  the adiabatic vacuum (see also Refs. \cite{c1,h1}).
Supposing $H_I(t)$ is small with respect to the free Hamiltonian, we can treat the expectation value perturbatively. To achieve  this, we need to know all possible contractions 
\be
G^{\pm\pm}(x, y) = \avg{{\rm T}_\mathcal{C}{\phi_\pm(x) \phi_\pm(y)}},
\ee
where $\phi(x)$ generically denotes a scalar field. 
The Green's functions can be expressed more explicitly, through the Heaviside $\theta(x)$ 
\begin{align*}
G^{-+}_{\phi}(x,y) &= \langle\phi(x)\phi(y)\rangle,\\
G^{+-}_{\phi}(x,y)&= \langle\phi(y)\phi(x)\rangle,\\
G^{++}_{\phi}(x,y) &=  \theta(x^0-y^0)G^{-+}_{\phi}(x,y)+ \theta(y^0-x^0)G^{+-}_{\phi}(x,y), \\
G^{--}_{\phi}(x,y) &=\theta(x^0-y^0)G^{+-}_{\phi}(x,y)+ \theta(y^0-x^0)G^{-+}_{\phi}(x,y),
\end{align*}
which  satisfy the  simple relation
\begin{equation}\label{eq:Grel}
G^{++}_{\phi}(x,y) + G^{--}_{\phi}(x,y) = G^{+-}_{\phi}(x,y) + G^{-+}_{\phi}(x,y).
\end{equation}
In our analysis, we will suppose that the initial density matrix is simply the free field vacuum state and  because of the spatial invariance of this state, we can Fourier transform the Green's functions (for instance in Minkowski space-time)
\be
G^{+-}(x,y) = \langle\phi(y)\phi(x)\rangle = \int \frac{{\rm d}^3k}{(2\pi)^3} e^{-i\mathbf{k}\cdot(\mathbf{x}-\mathbf{y})}\quadra{ \frac{1}{2\omega_k} e^{i\omega_k\cdot(x^0-y^0)}},\,\,\,\omega_k = \sqrt{\mathbf{k}^2 + m^2}.
\ee
The same computations can be done analogously for  $G^{-+}(x,y) $. From these expressions we recognize the Fourier modes
\begin{align}
G^{-+}(\mathbf{k}, x_0, y_0) &=\frac{1}{2\omega_k} e^{-i\omega_k\cdot(x^0-y^0)}, \\
G^{+-}(\mathbf{k}, x_0, y_0) &= \frac{1}{2\omega_k} e^{i\omega_k\cdot(x^0-y^0)}.
\end{align}
\section{Renormalization of composite operators  in a Minkowski time-dependent background}
The in-in  formalism, briefly summarized in the previous section,  will now be applied to a simple, yet illustrative example. Let us consider a field theory containing a light field $\phi(x)$ and a heavy field $\chi(x)$ with Lagrangian density 
\begin{equation}\label{eq:lagrangianaprincipale}
\mathcal{L}[\phi, \chi] = {\frac{1}{2}\partial_\mu\phi\partial^\mu\phi - \frac{1}{2}m^2\phi^2 + \frac{1}{2}\partial_\mu\chi\partial^\mu\chi - \frac{1}{2}M^2\chi^2 - \frac{g^2}{2}\phi^2\chi^2}.
\end{equation}
Translational invariance is explicitly broken  by imposing an initial  time $t_\text{in}$ in the action

\begin{equation}\label{eq:action}
S[\phi, \chi] = \int_{t_\text{in}}^\infty \d \tau \int \d^3 x\,\, \mathcal{L}[\phi, \chi].
\end{equation}
According to the in-in formalism, we have to double the degrees of freedom for both $\phi(x)$ and $\chi(x)$.
Fields $\phi_+$, $\chi_+$ propagate in the first part of the contour and $\phi_-$, $\chi_-$ in the last part.
The in-in action becomes
\be
S[\phi_+, \phi_-, \chi_+, \chi_-] = \int_{t_\text{in}}^\infty {\rm d} \tau \int {\rm d}^3 x\,\, {\mathcal{L}[\phi_+, \chi_+]-\mathcal{L}[\phi_-, \chi_-]}.
\ee
The Feynman rules can be read directly from the Lagrangian density
and in what follows we will introduce our conventions  to represent fields,  propagators, and vertices, as follows

\begin{center}{\it Fields}\end{center}
\begin{minipage}{.25\textwidth}
	\centering
	\includegraphics[width=.50\textwidth]{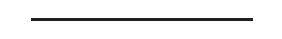}
\end{minipage}
\begin{minipage}{.25\textwidth}
	$= \phi_+$,
\end{minipage}
\vspace{2mm}
\begin{minipage}{.25\textwidth}
	\centering
	\includegraphics[width=.50\textwidth]{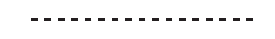}
\end{minipage}
\begin{minipage}{.25\textwidth}
	$= \phi_-$,
\end{minipage}\\
\begin{minipage}{.25\textwidth}
	\centering
	\includegraphics[width=.50\textwidth]{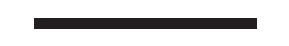}
\end{minipage}
\begin{minipage}{.25\textwidth}
	$= \chi_+$,
\end{minipage}
\vspace{2mm}
\begin{minipage}{.25\textwidth}
	\centering
	\includegraphics[width=.50\textwidth]{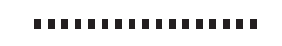}
\end{minipage}
\begin{minipage}{.25\textwidth}
	$= \chi_-$.
\end{minipage} 
\begin{center}{\it Propagators}\end{center}
\begin{minipage}{.25\textwidth}
	\centering
	\includegraphics[width=.50\textwidth]{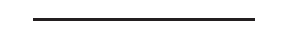}
\end{minipage}
\begin{minipage}{.25\textwidth}
	$= G^{++}_\phi(\mathbf{k}, x_0, y_0)$,
\end{minipage}
\vspace{2mm}
\begin{minipage}{.25\textwidth}
	\centering
	\includegraphics[width=.50\textwidth]{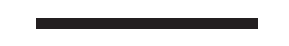}
\end{minipage}
\begin{minipage}{.25\textwidth}
	$= G^{++}_\chi(\mathbf{k}, x_0, y_0)$,
\end{minipage} 
\begin{minipage}{.25\textwidth}
	\centering
	\includegraphics[width=.50\textwidth]{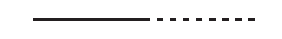}
\end{minipage}
\begin{minipage}{.25\textwidth}
	$= G^{+-}_\phi(\mathbf{k}, x_0, y_0)$,
\end{minipage}
\vspace{2mm}
\begin{minipage}{.25\textwidth}
	\centering
	\includegraphics[width=.50\textwidth]{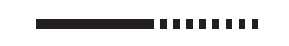}
\end{minipage}
\begin{minipage}{.25\textwidth}
	$= G^{+-}_\chi(\mathbf{k}, x_0, y_0)$,
\end{minipage}\\
\begin{minipage}{.25\textwidth}
	\centering
	\includegraphics[width=.50\textwidth]{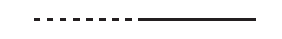}
\end{minipage}
\begin{minipage}{.25\textwidth}
	$= G^{-+}_\phi(\mathbf{k}, x_0, y_0)$,
\end{minipage}
\vspace{2mm}
\begin{minipage}{.25\textwidth}
	\centering
	\includegraphics[width=.50\textwidth]{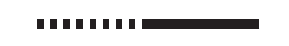}
\end{minipage}
\begin{minipage}{.25\textwidth}
	$= G^{-+}_\chi(\mathbf{k}, x_0, y_0)$,
\end{minipage}\\
\begin{minipage}{.25\textwidth}
	\centering
	\includegraphics[width=.50\textwidth]{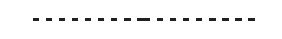}
\end{minipage}
\begin{minipage}{.25\textwidth}
	$= G^{--}_\phi(\mathbf{k}, x_0, y_0)$,
\end{minipage}
\vspace{2mm}
\begin{minipage}{.25\textwidth}
	\centering
	\includegraphics[width=.50\textwidth]{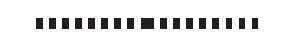}
\end{minipage}
\begin{minipage}{.25\textwidth}
	$= G^{--}_\chi(\mathbf{k}, x_0, y_0)$.
\end{minipage}
\begin{center}{\it Vertices}\end{center}
\begin{minipage}{.25\textwidth}
	\centering
	\includegraphics[width=.50\textwidth]{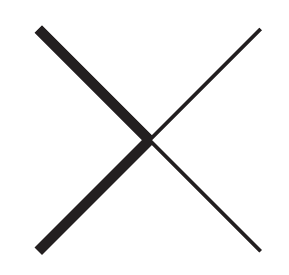}
\end{minipage}
\begin{minipage}{.25\textwidth}
	$= {-\frac{ig^2}{2}}$,
\end{minipage}
\vspace{2mm}
\begin{minipage}{.25\textwidth}
	\centering
	\includegraphics[width=.50\textwidth]{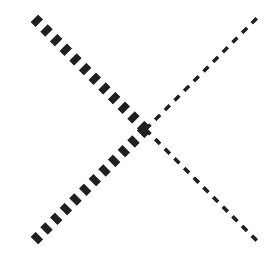}
\end{minipage}
\begin{minipage}{.32\textwidth}
	$= {\frac{ig^2}{2}}$.
\end{minipage}
Feynman rules for vertices are given without taking into account the symmetric factor and they should be properly added when computing
loops.

In quantum field theory one frequently encounters  products of fields, such  as $\phi(x)\phi(y)$. These products are called composite operators and  are singular at short distances, i.e. when $x \to y$. This is the limit probed by high energies and large momentum transfers.
Such operators usually appear in the Lagrange function and in relevant  operators such as the stress-energy tensor. The latter case is of particular interest because the matrix elements can be measured and therefore must be finite. It is therefore physically important to construct renormalized composite operators. Moreover, composite operators are a necessary ingredient for the operator product expansion, which is an essential tool in quantum field theory. It is used, for example, in the analysis of large momentum transfer inelastic scattering processes. Since the relativistic field theory is singular in the short-distance limit, the composite operators must be carefully defined in a regulated theory and divergent quantities must be subtracted to form renormalized operators. The purpose of this section is to illustrate the general character of composite operators in a time-dependent background and operator mixing through the toy Lagrangian we have introduced above.
In particular, we wish to show how the composite operators
 $\phi^2({\bf x},t)$, $\chi^2({\bf x},t)$, $\phi({\bf x},t)\chi({\bf x},t)$  mix with each other at first order in $g^2$ and what are the implications of being in a time-dependent background. \\
 
The expansion of the composite operator $\phi^2({\bf x},t)$ can be found considering the sum of all possible connected Green's functions of the form
\be
 \avg{\phi_i(t)\phi_j(t) \chi_k(t_1) \chi_l(t_2)}_c,\qquad i,j,k,l\in\{+, -\}.
\ee
For each Green's function, one needs to consider only four Feynman diagrams to the order $g^2$. For instance, for the correlator $\avg{\phi_+^2(t) \chi_+(t_1) \chi_+(t_2)}_c$, two diagrams are given in Fig. \ref{fig:Feynman}. The remaining ones, $A_2$ and $B_2$, are obtained by exchanging $\mathbf{p_1}$ and $\mathbf{p_2}$.
\begin{figure}[t]
\centering
\includegraphics[scale=0.85]{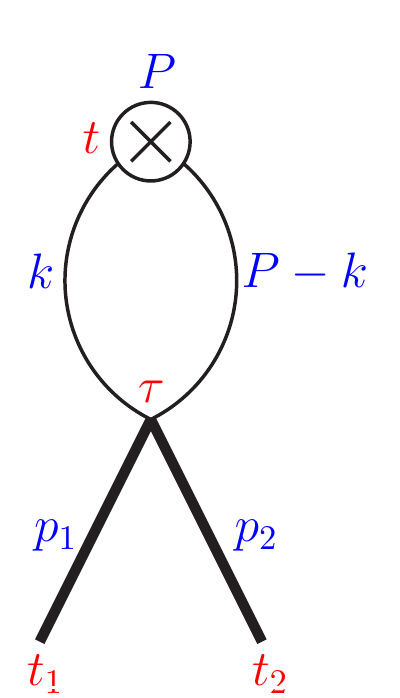}
\hspace{3.5cm}
\includegraphics[scale=0.85]{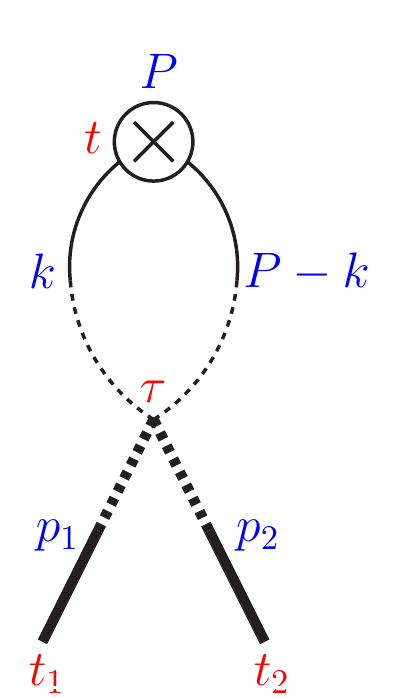}
\caption{\label{fig:Feynman}\emph{{\small
Feynman diagrams $A_1$ (left) and $B_1$ (right) for the renormalization of the operator $\phi^2$. The composite operators $\phi_i\phi_j, \, i,j\in\{+,-\}$, are conventionally denoted by a wheel cross vertex $\otimes$. }}} 
\end{figure}
\noindent
 These contributions in momentum space are
	\begin{eqnarray}
	({\it A}_1 + {\it A}_2)&=&2\,{\frac{-ig^2}{2}}\int \frac{{\rm d}^3k}{(2\pi)^3}\int_{t_{\text{in}}}^\infty {\rm d} \tau \, G^{++}_\phi(\mathbf{k}, t, \tau)G^{++}_\phi(\mathbf{P-k}, t, \tau)G^{++}_\chi(\mathbf{p_1}, \tau, t_1)G^{++}_\chi(\mathbf{p_2}, \tau, t_2)\nonumber  \\ &+& (\mathbf{p_1}\leftrightarrow \mathbf{p_2}),
	\end{eqnarray} 
	and
	\begin{eqnarray}
	({\it B}_1 +{\it B}_2)&=&2\,{\frac{ig^2}{2}}\int \frac{{\rm d}^3k}{(2\pi)^3}\int_{t_{\text{in}}}^\infty {\rm d} \tau \, G^{+-}_\phi(\mathbf{k}, t, \tau)G^{+-}_\phi(\mathbf{P-k}, t, \tau)G^{-+}_\chi(\mathbf{p_1}, \tau, t_1)G^{-+}_\chi(\mathbf{p_2}, \tau, t_2) \nonumber\\&+& (\mathbf{p_1}\leftrightarrow \mathbf{p_2}).
	\end{eqnarray}
The two integrals involve only exponential functions and can then be expressed in a closed form. The only difficulties are the Heaviside functions inside the propagators $G^{++}$ and $G^{--}$. This forces us to consider separate cases, each corresponding to a different temporal ordering of $t$, $t_1$, and $t_2$.

We will present the  explicit calculation only for $t<t_1<t_2$.  The other orderings give exactly the same result  as expected on physical grounds.  Diagrams $A_1$ and $A_2$ give

\begin{eqnarray}
({\it A}_1 + {\it A}_2)&=&2\,{\frac{-ig^2}{2}}\int \frac{{\rm d}^3k}{(2\pi)^3}\int_{t_{\text{in}}}^\infty {\rm d} \tau \, G^{++}_\phi(\mathbf{k}, t, \tau)G^{++}_\phi(\mathbf{P-k}, t, \tau)G^{++}_\chi(\mathbf{p_1}, \tau, t_1)G^{++}_\chi(\mathbf{p_2}, \tau, t_2) + (\mathbf{p_1}\leftrightarrow \mathbf{p_2})\nonumber\\
&=& -ig^2\int  \frac{{\rm d}^3k}{(2\pi)^3}\frac{1}{16\omega_k\omega_{P-k}\omega_{p_1}\omega_{p_2}}\bigg\{\int_{t_\text{in}}^t {\rm d} \tau \, e^{-i\omega_k(t-\tau)}e^{-i\omega_{P-k}(t-\tau)}e^{+i\omega_{p_1}(\tau-t_1)}e^{+i\omega_{p_2}(\tau-t_2)}\nonumber \\
&+&\int_{t_2}^{\infty} {\rm d} \tau \, e^{+i\omega_k(t-\tau)}e^{+i\omega_{P-k}(t-\tau)}e^{-i\omega_{p_1}(\tau-t_1)}e^{-i\omega_{p_2}(\tau-t_2)\nonumber}\\
&+&\int_{t_1}^{t_2} {\rm d} \tau \, e^{+i\omega_k(t-\tau)}e^{+i\omega_{P-k}(t-\tau)}e^{-i\omega_{p_1}(\tau-t_1)}e^{+i\omega_{p_2}(\tau-t_2)}\nonumber \\
&+&\int_{t_\text{in}}^{t_1} {\rm d} \tau \, e^{+i\omega_k(t-\tau)}e^{+i\omega_{P-k}(t-\tau)}e^{+i\omega_{p_1}(\tau-t_1)}e^{+i\omega_{p_2}(\tau-t_2)}\nonumber \\
&-&\int_{t_\text{in}}^{t} {\rm d} \tau \, e^{+i\omega_k(t-\tau)}e^{+i\omega_{P-k}(t-\tau)}e^{+i\omega_{p_1}(\tau-t_1)}e^{+i\omega_{p_2}(\tau-t_2)} \bigg\} + (\omega_{p_1}\leftrightarrow \omega_{p_2}),
\end{eqnarray}
while diagrams $B_1$ and $B_2$ give
\begin{eqnarray}
({\it B}_1 + {\it B}_2)&=&2\,{\frac{ig^2}{2}}\int \frac{{\rm d}^3k}{(2\pi)^3}\int_{t_{\text{in}}}^\infty {\rm d} \tau \, G^{+-}_\phi(\mathbf{k}, t, \tau)G^{+-}_\phi(\mathbf{P-k}, t, \tau)G^{-+}_\chi(\mathbf{p_1}, \tau, t_1)G^{-+}_\chi(\mathbf{p_2}, \tau, t_2)+ (\mathbf{p_1}\leftrightarrow \mathbf{p_2})\nonumber \\
&=& ig^2\int  \frac{{\rm d}^3k}{(2\pi)^3}\frac{1}{16\omega_k\omega_{P-k}\omega_{p_1}\omega_{p_2}}\int_{t_\text{in}}^\infty {\rm d} \tau \, e^{+i\omega_k(t-\tau)}e^{+i\omega_{P-k}(t-\tau)}e^{-i\omega_{p_1}(\tau-t_1)}e^{-i\omega_{p_2}(\tau-t_2)}\nonumber \\
&+& (\omega_{p_1}\leftrightarrow \omega_{p_2}).
\end{eqnarray} 
The short-distance expansion is done in the large momentum $\mathbf{k}$ limit which allows us to take \mbox{$\omega_{P-k} \sim \omega_k$}. Adding up all the contributions we obtain
\begin{eqnarray}\label{eq:A+B}
({\it A}_1+{\it A}_2+{\it B}_1+{\it B}_2)&=&\int  \frac{{\rm d}^3k}{(2\pi)^3}\frac{g^2}{32 \omega_k^3 \omega_{p_1}\omega_{p_2}}\bigg\{e^{-i \quadra{2 \omega_k(t-t_\text{in})+\omega_{p_1}(t_1-t_\text{in})+\omega_{p_2}(t_2-t_\text{in})}}\nonumber\\
&+&e^{i\quadra{2 \omega_k(t-t_\text{in})+\omega_{p_1}(t_1-t_\text{in})+\omega_{p_2}(t_2-t_\text{in})}}\nonumber \\
&-& 2e^{i\omega_{p_1}(t-t_1)}e^{i\omega_{p_2}(t-t_2)} \bigg\}+ (\omega_{p_1}\leftrightarrow \omega_{p_2}).
\end{eqnarray} 
Note that (\ref{eq:A+B}) is just one of the terms which would contribute to the final four-point correlator and needs not to be self-conjugate. The result can be written in terms of Green's functions that involve only $\chi$-fields
\begin{eqnarray}
\avg{\phi_+^2(t) \chi_+(t_1) \chi_+(t_2)}_c&=& \int\frac{{\rm d}^3k}{(2\pi)^3}\frac{g^2}{8 \omega_k^3}\bigg\{e^{-2 i \omega_k(t-t_\text{in})}\avg{\chi_+^2(t_\text{in}) \chi_+(t_1) \chi_+(t_2)}_c
\nonumber\\
&+&e^{2 i \omega_k(t-t_\text{in})}\avg{\chi_-^2(t_\text{in}) \chi_+(t_1) \chi_+(t_2)}_c\nonumber\\
&-&2 \avg{\chi_+^2(t) \chi_+(t_1) \chi_+(t_2)}_c\bigg\}.
\end{eqnarray}
Notice that on the right-hand side the connected contributions are not zero as  two fields (out of four) are computed at the same point. Therefore, to the zeroth order in $g^2$  the four-point functions have  connected parts. 
The other Green's functions are computed similarly and can be expressed compactly in the form
\begin{eqnarray}
\avg{\phi_i(t)\phi_j(t)\chi_k(t_1)\chi_l(t_2)}_c &=& \int \frac{{\rm d}^3k}{(2\pi)^3}\frac{g^2}{8\omega_k^3}\bigg\{ e^{-2i\omega_k(t-t_\text{in})}\avg{\chi_+^2(t_{\rm in})\chi_k(t_1)\chi_l(t_2)}_c\nonumber\\
&+& e^{2i\omega_k(t-t_\text{in})}\avg{\chi_-^2(t_{\rm in})\chi_k(t_1)\chi_l(t_2)}_c\nonumber \\
&-&2\avg{\chi_i(t)\chi_j(t)\chi_k(t_1)\chi_l(t_2)}_c\bigg\}, \qquad i,j,k,l\in\{+,-\}.
\end{eqnarray}
The sum of all Green's functions gives the final self-conjugated four-point correlator
\begin{align}
&\bigg<\left(\frac{\phi_+(t)+\phi_-(t)}{2}\right)^2\left(\frac{\chi_+(t_1)+\chi_-(t_1)}{2}\right)\left(\frac{\chi_+(t_2)+\chi_-(t_2)}{2}\right)\bigg>_c\nonumber \\
&\qquad\qquad= \int\frac{{\rm d}^3k}{(2\pi)^3} \frac{g^2}{8\omega_k^3}\bigg\{ e^{-2i\omega_k(t-t_\text{in})}\avg{\chi_+^2(t_{\rm in})\left(\frac{\chi_+(t_1)+\chi_-(t_1)}{2}\right)\left(\frac{\chi_+(t_2)+\chi_-(t_2)}{2}\right)}_c\nonumber\\
&\qquad\qquad+ e^{2i\omega_k(t-t_\text{in})}\avg{\chi_-^2(t_{\rm in})\left(\frac{\chi_+(t_1)+\chi_-(t_1)}{2}\right)\left(\frac{\chi_+(t_2)+\chi_-(t_2)}{2}\right)}_c\nonumber \\
&\qquad\qquad-2\avg{\left(\frac{\chi_+(t)+\chi_-(t)}{2}\right)^2\left(\frac{\chi_+(t_1)+\chi_-(t_1)}{2}\right)\left(\frac{\chi_+(t_2)+\chi_-(t_2)}{2}\right)}_c\bigg\}.
\end{align}
At this point, we should remember that the fields $+$ and $-$ were added to account properly for the time evolution. 
In order to go back to the    physical fields $\phi$ and $\chi$ we need to  set $\phi_+=\phi_-=\phi$ and $\chi_+=\chi_-=\chi$. We  obtain 
for the connected part of the expectation values
\begin{eqnarray}
\avg{\phi^2(t) \chi(t_1) \chi(t_2)}_c &=&  \frac{g^2}{4}\bigg\{\int \frac{{\rm d}^3k}{(2\pi)^3}\frac{\cos(2\omega_k(t-t_\text{in}))}{\omega_k^3}\avg{\chi^2(t_\text{in}) \chi(t_1) \chi(t_2)}_c -  \int \frac{{\rm d}^3k}{(2\pi)^3}\frac{1}{\omega_k^3}\avg{\chi^2(t) \chi(t_1) \chi(t_2)}_c\bigg\} \nonumber\\
&=&  \frac{g^2}{4}\bigg\{K^\phi(t-t_\text{in})\avg{\chi^2(t_\text{in}) \chi(t_1) \chi(t_2)}_c - K^\phi(0)\avg{\chi^2(t) \chi(t_1) \chi(t_2)}_c\bigg\}.
\end{eqnarray}
This is equivalent to the operatorial relation for the properly normalized quantities at the scale $Q$
\be
\label{aqw}
\phi_Q^2(t) =\phi_0^2(t)+  \frac{g^2}{4}\left[K^\phi(t-t_\text{in})\chi_0^2(t_\text{in}) - K^\phi(0)\chi_0^2(t)\right]+{\rm counter\text{-}terms},
\ee
where the subscript $_0$ indicates bare quantities. 
The memory kernel $K^\phi$ is defined as
\begin{equation}\label{eq:Kphi}
K^\phi(t) = \int \frac{{\rm d}^3k}{(2\pi)^3}\frac{\cos(2\omega^\phi_{k}t)}{(\omega^\phi_{k})^3},\,\,\omega_k^\phi=\sqrt{{\bf k}^2+m^2}.
\end{equation}
Notice that the mixing vanishes at the initial time $t=t_\text{in}$.
Our results show that  composite operators are mixed with each other in time-dependent backgrounds. This does not come as a surprise.
Composite operator renormalization
induces a mixing among  composite operators which, for instance, 
make a definition of number densities in an interacting theory quite
cumbersome.

What is less trivial is that  renormalized fields are composed of two types of terms: one is the standard local term and    the 
time-dependence appears only in the  fields, the other  is non-local because the fields are multiplied by a function which depends on the elapsed time. This introduces a memory effect once the initial conditions of the problem are set. This is typical of non-equilibrium systems. Let us consider the mixing in Eq. (\ref{aqw}).
The  non-local piece  is expressed in terms of   a memory kernel 
\be
K^\phi(t-t_\text{in}) = \int \frac{{\rm d}^3 k}{(2\pi)^3} \, \frac{\cos\quadra{2 \omega^\phi_k (t-t_\text{in})}}{(\omega^{\phi}_k)^3}.
\ee
The kernel can be rewritten as a function of the new variable $z =m(t-t_\text{in})$ as 
\be
K^\phi(t-t_\text{in}) = \frac{1}{2\pi^2} \int_1^\infty{\rm d} k \sqrt{k^2-1}\, \frac{\cos{2k z}}{k^2},
\ee
and from this expression we can express it in terms of a Meijer G-function. Indeed, the cosine function is given by
\be
\cos(2\sqrt{x}z)=\sqrt{\pi}\,
\MeijerG{}{}{0}{\frac{1}{2}}{xz^2}.
\ee
Once inserted in the memory kernel we obtain
\be
K^\phi(t-t_\text{in}) = \frac{\sqrt{\pi}}{4\pi^2} \int_1^\infty {\rm d} x\,x^{-3/2}\sqrt{x-1}\,
\MeijerG{}{}{0}{\frac{1}{2}}{xz^2}=
\frac{1}{8\pi}\,
\MeijerG{}{\frac{3}{2}}{0,0}{\frac{1}{2}}{z^2}.
\ee
For large $m(t-t_\text{in})$ the memory kernel has the asymptotic behavior
\be
\label{memory}
K^\phi(t-t_\text{in}) \sim \frac{\cos\quadra{2m(t-t_\text{in})+\frac{3}{4}\pi}}{\quadra{4\pi m(t-t_\text{in})}^\frac{3}{2}},
\ee
and vanishes as  $(mt)^{-3/2}$ in the limit $m(t-t_\text{in})\to \infty$. On the contrary,  near the initial time $m(t-t_\text{in}) \ll 1$, the memory kernel has a logarithmic behavior

\be
K^\phi(t-t_\text{in}) \sim -\frac{1+\gamma + \ln[ m(t-t_\text{in})]}{2\pi^2}\sim -\frac{ \ln[ m(t-t_\text{in})]}{2\pi^2}.
\ee
Let us extract the divergent part of  $K^\phi(0)$ by regulating the theory with a momentum cutoff $\Lambda$.
\begin{eqnarray}\label{eq:k(0)}
K^\phi(0)&=& \int^\Lambda \frac{{\rm d}^3 k}{(2\pi)^3} \frac{1}{(\omega^{\phi}_k)^3}=\frac{1}{2\pi^2}\Bigg({-\frac{1}{\sqrt{1+\frac{m^2}{\Lambda^2}}}+\ln{\Lambda\sqrt{1+\frac{m^2}{\Lambda^2}}}}\Bigg)\nonumber\\
&=&\frac{1}{2\pi^2}\left(-1+\ln{2\Lambda}\right)+{\cal O}\left(\frac{1}{\Lambda}\right) \sim \frac{\ln\Lambda^2}{4\pi^2}.
\end{eqnarray}
By removing such a logarithmic divergence by a counter-term and imposing the renormalization condition that no mixing is present at the  scale
$Q^2=M^2$  at $(t-t_\text{in})\gg m^{-1}$,  we find at ${\cal O}(g^2)$ 
\be\label{eq:phiOPE}
\phi_Q^2(t)=\phi_M^2(t)-\frac{g^2}{8\pi^2}\ln\left(\frac{M^2}{Q^2}\right)\,\chi_M^2(t)+\frac{g^2}{4}K^\phi(t-t_{\text{in}})\chi_M^2(t_\text{in}).
\ee
A completely analogous computation gives  at $(t-t_\text{in})\gg M^{-1}$
\be\label{eq:phiOPE}
\chi_Q^2(t)=\chi_M^2(t)-\frac{g^2}{8\pi^2}\ln\left(\frac{M^2}{Q^2}\right)\,\phi_M^2(t)+\frac{g^2}{4}K^\chi(t-t_{\text{in}})\phi_M^2(t_\text{in}),
\ee
where 
\begin{equation}\label{eq:Kchi}
K^\chi(t) = \int \frac{{\rm d}^3k}{(2\pi)^3}\frac{\cos(2\omega^\chi_{k}t)}{(\omega^\chi_{k})^3},\,\,\omega_k^\chi=\sqrt{{\bf k}^2+M^2},
\end{equation}
and
\be\label{eq:phiOPE}
(\phi(t)\chi(t))_Q=(\phi(t)\chi(t))_M-\frac{g^2}{8\pi^2}
\ln\left(\frac{M^2}{Q^2}\right)\,
(\phi(t)\chi(t))_M+\frac{g^2}{2}J(t-t_{\text{in}})
(\phi(t_\text{in})\chi(t_\text{in}))_M,
\ee
with
\begin{equation}\label{eq:J}
J(t) = \int \frac{{\rm d}^3k}{(2\pi)^3}\frac{\cos((\omega^\phi_{k}+\omega^\chi_{k})t)}{\omega^\phi_{k}\omega^\chi_{k}(\omega^\phi_{k}+\omega^\chi_{k})}.
\end{equation}
From Eq. (\ref{memory}) we see that it takes a typical scale $t\sim m^{-1}$ in order for the memory effects to be negligible. In particular, there might be a hierarchy of time scales for the memory effects if $M\gg m$. This difference in the non-local kernel makes the diagonalization of the composite operators more difficult and might play an important role in some cosmological phenomena.

To interpret physically these results, one has to remember that in the in-in   formalism we have prepared the system at $t_\text{in}$ in a free-vacuum state $\rho_\text{in}$ and fields start evolving from this initial time. An equivalent scenario can be constructed by supposing that the system evolves freely before $t_\text{in}$ according to the free Lagrangian $\mathcal{L}_0$. The initial state can be arranged so as to obtain the exact density state $\rho_\text{in}$ at $t_\text{in}$. At this time the interaction term $\mathcal{L_\text{int}}=-(g^2/2)\phi^2\chi^2$ is introduced suddenly and  $\rho_\text{in}$ is no longer an eigenstate of the system. This perspective can be represented mathematically through a Heaviside function $\theta(t-t_\text{in})$ which multiplies the interaction term
\be
\label{actionin}
S[\phi, \chi] = \int_{-\infty}^{\infty} {\rm d} \tau \int {\rm d}^3 \mathbf{x}\,\, \big\{\mathcal{L}_0[\phi, \chi]+\theta(t-t_\text{in})\mathcal{L_\text{int}}[\phi, \chi] \big\}.
\ee
From this point of view,  the results obtained above indicate that, once the interaction is introduced in the system, a non-local term appears in the expression of the normalized composite operators. This term has an amplitude which decreases faster than $[m(t-t_{\rm in})]^{-3/2}$ and disappears at $t\to \infty$. 
Using this interpretation, we can physically interpret the appearance of the non-local term in the composite operator renormalization 
as a consequence of 
introducing a finite energy density into the system at the time $t_{\rm in}$. This implies that the modes of the fields may be excited
and their effect is to introduce a non-local mixing between the composite operators, which dies off as time goes by. 
Had we tuned on the interaction in a infinitely  adiabatic way,  the non-local term would  not be present because the  mode excitation would have been suppressed.  At any rate, what is interesting is that, once one sets the system at some  $t_\text{in}$,  the  non-local
effects have a  universal time-dependence characterized by a power-law decay $\sim t^{-3/2}$. The same memory kernel appears in the study of  the effects of the heavy field
on the dynamics of the light field by analyzing the equation of motion for the expectation
value of the light background field \cite{holman}. There too new non-local terms appear which cannot arise from a local action
of an effective field theory in terms of the light field, though they disappear in the adiabatic
limit.

To support the interpretation of our results, based on the excitation of the field modes when the interaction between the $\phi$- and $\chi$-fields is switched on, let us consider the solution of the equation for the $\phi$-field with the action (\ref{actionin})

\begin{eqnarray}
\phi({\bf k},t)&=& \frac{a_{\bf k}}{\sqrt{2 \omega_k^\phi}} \,e^{-i \omega_k^\phi t}+{\rm h.c.}\,\,\,(t< t_{\rm in}),\nonumber\\
\phi({\bf k},t)&=& \frac{\alpha_{\bf k}}{\sqrt{2 \Omega_k^\phi}}\,e^{-i \Omega_k^\phi t}+{\rm h.c.} \,\,\,(t> t_{\rm in}),\nonumber\\
\Omega_k^\phi&=&\left({\bf k}^2+m^2+g^2\chi^2(t_{\rm in})\right)^{1/2},\nonumber\\
\nonumber\\
\alpha_\mathbf{k}&=&A^*_k a_\mathbf{k} - B^*_k a_\mathbf{k}^\dagger, \nonumber\\
A_k&=&+\frac{\left(\Omega_k^\phi-\omega_k^\phi\right)}{2\sqrt{\omega_k^\phi \Omega_k^\phi}}\,{\rm exp}\left[-i\left(\Omega_k^\phi-\omega_k^\phi\right)t_{\rm in}\right],\nonumber\\
B_k&=&-\frac{\left(\Omega_k^\phi+\omega_k^\phi\right)}{2\sqrt{\omega_k^\phi \Omega_k^\phi}}\,{\rm exp}\left[-i\left(\Omega_k^\phi+\omega_k^\phi\right)t_{\rm in}\right],
\end{eqnarray}
where we have made the simplifying approximation that  $\chi$-field is very slowly   changing with \mbox{time, as we are} interested in the UV (equal point) limit of the composite operator $\phi^2({\bf x},t)$. In Fourier space it is \mbox{given by}

\begin{align}
\phi^2(\mathbf{k}, t) &=\frac{1}{2\Omega_k^\phi}\Bigg\{\frac{a_\mathbf{k}a_\mathbf{k}^\dagger + a_\mathbf{k}^\dagger a_\mathbf{k}}{2\omega_k^\phi\Omega_k^\phi}\bigg[(\omega_k^\phi)^2+(\Omega_k^\phi)^2 + \Big((\Omega_k^\phi)^2-(\omega_k^\phi)^2\Big)\cos\big(2\Omega_k^\phi(t-t_\text{in})\big)\bigg] \nonumber\\
&+\frac{a_\mathbf{k}^2}{\omega_k^\phi\Omega_k^\phi}\bigg[e^{-2 i \omega_k^\phi t_\text{in}}\Big(\Omega_k^\phi\cos\big(\Omega_k^\phi(t-t_\text{in})\big)+i\omega_k^\phi\sin\big(\Omega_k^\phi(t-t_\text{in})\big)\Big)^2\bigg]\nonumber\\
&+\frac{(a_\mathbf{k}^\dagger)^2}{\omega_k^\phi\Omega_k^\phi}\bigg[e^{2 i \omega_k^\phi t_\text{in}}\Big(\Omega_k^\phi\cos\big(\Omega_k^\phi(t-t_\text{in})\big)-i\omega_k^\phi\sin\big(\Omega_k^\phi(t-t_\text{in})\big)\Big)^2\bigg]\Bigg\}.
\end{align}
By computing the correlator $\big< \phi^2(t)\chi(t_1)\chi(t_2)\big>$, it is easy to see that the last term of the first line  
reproduces the  non-local memory kernel 
\be\label{trick}
\Big<\phi^2(t)\chi(t_1)\chi(t_2)\Big>\supset
\frac{g^2}{4}\int \frac{{\rm d}^3 k}{(2\pi)^3} \, \frac{\cos\quadra{2 \omega^\phi_k (t-t_\text{in})}}{(\omega^{\phi}_k)^3}
\Big<\chi^2(t_{\rm in})\chi(t_1)\chi(t_2)\Big>.
\ee
 This is due to the fact that at $t>t_{\rm in}$ negative frequency modes with amplitude proportional to $B_k$ appear and they
are to be interpreted as the excited states compared to the initial vacuum at $t<t_{\rm in}$
\be
\Big< 0\left| \alpha^\dagger_\mathbf{k}\alpha_\mathbf{k}\right|0\Big>=\left|B_k\right|^2. 
\ee
\section{Renormalization of composite operators  in a de Sitter time-dependent background}
In the previous section we have seen how the composite operators $\phi^2$, $\chi^2$, and $\phi\chi$ behave under renormalization in a time-dependent Minkowski space. We are now in the  position to apply the same procedure in de Sitter space described by the metric
\be
\d s^2=\frac{1}{(H\eta)^2}\left(\d \eta^2-\d \vec{x}^2\right),
\ee
where $\eta$ indicates the conformal time and $H$ is the constant Hubble rate. The Lagrangian density is given by 
\begin{equation}\label{eq:lagrangianaprincipaledS}
\mathcal{L}[\phi, \chi] =\left( \frac{1}{H\eta}\right)^4\left(\frac{1}{2}\partial_\mu\phi\partial^\mu\phi - \frac{1}{2}m^2\phi^2 + \frac{1}{2}\partial_\mu\chi\partial^\mu\chi - \frac{1}{2}M^2\chi^2 - \frac{g^2}{2}\phi^2\chi^2\right).
\end{equation}
We have not been able to perform an analytical computation of the renormalization for massive scalar fields. 
To simplify the problem we have therefore  considered the case in which both  fields $\phi$ and $\chi$ have  negligible  masses. In such  a case the  propagators are  given in terms of the Hankel functions $H_\mu^{(1,2)}$
\begin{eqnarray}
G^{-+}_{\phi}(\mathbf{k}, \eta, \eta') &=&\frac{\pi H^2 H_{3/2}^{(1)}(-k\eta)H_{3/2}^{(2)}(-k\eta')}{4}(\eta\eta')^{3/2}, \nonumber\\
G^{+-}_{\phi}(\mathbf{k}, \eta, \eta') &=&\frac{\pi H^2 H_{3/2}^{(2)}(-k\eta)H_{3/2}^{(1)}(-k\eta')}{4}(\eta\eta')^{3/2}.
\end{eqnarray}
The advantage of using the massless  approximation is that the Hankel functions can be given explicitly. Indeed,  for $\mu=\frac{3}{2}$  the  Hankel functions are  given by
\begin{eqnarray}
H_{3/2}^{(1)}(z) &=&-\sqrt{\frac{2}{\pi z}}e^{iz}\left(1 - \frac{1}{iz}\right), \nonumber\\
H_{3/2}^{(2)}(z) &=&\left[{H}_{3/2}^{(1)}(z)\right]^*.
\end{eqnarray}
In what follows we will perform another approximation, that is we will  investigate the composite operator renormalization on super-Hubble scales for the two external legs  $p_1,p_2 \ll aH$. This allows us to simplify further the external propagators. Indeed,  in this regime, $z\ll 1$ for the external legs  and the Hankel functions can be approximated by the dominant term
\begin{eqnarray}
H_{3/2}^{(1)}(z) &=&-i\sqrt{\frac{2}{\pi}}\left(\frac{1}{z}\right)^\frac{3}{2},\nonumber \\
{H}_{3/2}^{(2)}(z) &=&\left[{H}_{3/2}^{(1)}(z)\right]^*.
\end{eqnarray}
The starting point is again  to compute perturbatively the Green's function 
 $\langle\phi_+^2(\eta) \chi_+(\eta_1) \chi_+(\eta_2)\rangle$ through the same diagrams of Fig. 2.
Before reporting on  the computation of the corresponding diagrams, we need to discuss how to regularize momentum integrals. In de Sitter space, we interpret $\Lambda$ as the physical cutoff at which the dynamics of some  heavy sector intervenes.  
Since  $\Lambda$ is a physical cutoff, it  regulates integrals over physical momenta $k_\text{phys}$. Since the  integrals are over comoving momenta $k = k_{\rm phys} a$, the cutoff becomes time-dependent and equal to $\Lambda a$. The theory can therefore be regulated by the replacement
\be
\int \d \tau \int \d^3 k\to \int \d \tau \int^{\Lambda a(\tau)}\d^3 k.
\ee
The corresponding regulated expressions are therefore

\begin{eqnarray}
(A_1+A_2)&=&-2ig^2\int_{\eta_{\text{in}}}^0 \d \tau \int^{\Lambda a(\tau)} \frac{\d^3k}{(2\pi)^3}\left(\frac{1}{\tau H}\right)^4G^{++}_\phi(\mathbf{k}, \eta, \tau)G^{++}_\phi(\mathbf{P-k}, \eta, \tau) G^{++}_\chi(\mathbf{p_1}, \tau, \eta_1)G^{++}_\chi(\mathbf{p_2}, \tau, \eta_2),\nonumber\\
&&
\end{eqnarray}
and

\begin{eqnarray}
(B_1+B_2)&=&2ig^2\int_{\eta_{\text{in}}}^0 \d \tau \int^{\Lambda a(\tau)} \frac{\d^3k}{(2\pi)^3}\left(\frac{1}{\tau H}\right)^4G^{+-}_\phi(\mathbf{k}, \eta, \tau)G^{+-}_\phi(\mathbf{P-k}, \eta, \tau) G^{-+}_\chi(\mathbf{p_1}, \tau, \eta_1)G^{-+}_\chi(\mathbf{p_2}, \tau, \eta_2).\nonumber\\
&&
\end{eqnarray}
It  is  useful to change the momentum integration variable in order to eliminate the temporal dependence of the momentum integration domain
\be
k \to z = \frac{k}{a(\tau)H}=-k\tau.
\ee
Now we are able to perform the same manipulations, as in the previous section. First, because of the Heaviside functions inside the propagators, we are obliged to consider different temporal orderings. Some cases are not treated, but can be easily recovered by exchanging $\eta_1\to \eta_2$ and $\mathbf{p_1}\to\mathbf{p_2}$. We focus only on the temporal ordering $\eta_{\rm in}<\eta<\eta_1<\eta_2$ as the others give similar results.  We are interested in  the  large $z$ (or equivalently large momenta), when the external fields are on super-Hubble scales. This justifies a first approximation in the propagators
\be
\mathbf{P}+\frac{\mathbf{z}}{\tau}\sim  \frac{\mathbf{z}}{\tau}\,\,{\rm for}\,\, z\gg P\tau
\ee
and the use of simplified Hankel functions  for the external legs.
Once all the contributions are summed, we obtain

\begin{eqnarray}
(A_1+A_2+B_1+B_2)&=&
\frac{g^2 H^4}{32p_1^3p_2^3}\int^{\frac{\Lambda}{H}} \frac{\d^3z}{(2\pi)^3}\frac{1}{z^6}
 \bigg\{\bigg[e^{-2iz\frac{\eta-\eta_\text{in}}{\eta_\text{in}}}\left(\frac{2z\eta}{\eta_\text{in}}-5i\right)(z+i)^2+e^{2iz\frac{\eta-\eta_\text{in}}{\eta_\text{in}}}\left(\frac{2z\eta}{\eta_\text{in}}+5i\right)(z-i)^2\bigg]\nonumber\\ 
 &-&4z(4+z^2)
 +4i \bigg[e^{-2iz}(z-i)^2\bigg(\text{Ei}(2iz)-\text{Ei}\left(\frac{2iz\eta}{\eta_\text{in}}\right)\bigg)\nonumber\\
 &-&e^{2iz}(z+i)^2\bigg(\text{Ei}(-2iz)-\text{Ei}\left(-\frac{2iz\eta }{\eta_\text{in}}\right)\bigg)\bigg]\bigg\},\nonumber\\
 &&
\end{eqnarray}
where Ei is the exponential integral function 
\be
\text{Ei}(ix) = -\int_x^\infty \d t \, \frac{\cos t}{t}+i\left[\frac{\pi}{2}-\int_x^\infty \d t\, \frac{\sin t}{t}\right].
\ee
We are now able to recognize the two-point Green's function for $\chi^2$
\be
\langle\chi_+^2(\eta) \chi_+(\eta_1) \chi_+(\eta_2)\rangle_c = \frac{H^4}{2p_1^3p_2^3},
\ee
which can be used to express the four-point function $\langle\phi_+^2(\eta) \chi_+(\eta_1) \chi_+(\eta_2)\rangle_c$ as 
\begin{eqnarray}
\langle\phi_+^2(\eta) \chi_+(\eta_1) \chi_+(\eta_2)\rangle_c&=&
\frac{g^2}{8}\avg{\chi_+^2(\eta) \chi_+(\eta_1) \chi_+(\eta_2)}_c\int^{\frac{\Lambda}{H}} \frac{\d^3z}{(2\pi)^3}\frac{1}{z^6}\bigg\{ -2z(4+z^2) \nonumber\\
&+&\Re\bigg[e^{2iz\left(\frac{\eta-\eta_\text{in}}{\eta_\text{in}}\right)}
\left(\frac{2z\eta}{\eta_\text{in}}+5i\right)(z-i)^2\bigg]\nonumber \\
&+& 4\Im \bigg[e^{2iz}(z+i)^2\bigg(\text{Ei}(-2iz)-\text{Ei}\left(-\frac{2iz\eta}{\eta_\text{in}}\right)\bigg)\bigg]\bigg\}.
\nonumber\\
&&
\end{eqnarray}
The dominant term is given by taking the first non-zero term of the Taylor expansion around $z=+\infty$
\begin{eqnarray}
\langle\phi_+^2(\eta) \chi_+(\eta_1) \chi_+(\eta_2)\rangle_c&=&
\frac{g^2}{8\pi^2}\langle\chi_+^2(\eta) \chi_+(\eta_1) \chi_+(\eta_2)\rangle_c\int^{\frac{\Lambda}{H}}
 \frac{\d z}{z^3}
\bigg\{ -z^2 +\cos\left(2z\left(\frac{\eta-\eta_\text{in}}{\eta_\text{in}}\right)\right)\left(z^2\frac{\eta}{\eta_\text{in}}-\frac{\eta_\text{in}}{\eta}\right)\bigg\}.\nonumber\\
&&
\end{eqnarray}
This expression can be integrated analytically and expressed as a function of the exponential integral and elementary functions
\begin{eqnarray}
\langle\phi_+^2(\eta) \chi_+(\eta_1) \chi_+(\eta_2)\rangle_c &=&
\frac{g^2}{16\pi^2}\langle\chi_+^2(\eta) \chi_+(\eta_1) \chi_+(\eta_2)\rangle_c\bigg\{-2\ln\left(\frac{\Lambda}{H}\right) \nonumber\\
&+&\frac{H^2}{\Lambda^2}\left(\frac{\eta_\text{in}}{\eta}\cos\left[2\frac{\Lambda}{H}\left(\frac{\eta-\eta_\text{in}}{\eta_\text{in}}\right)\right]+2\frac{\Lambda}{H}\left(\frac{\eta-\eta_\text{in}}{\eta}\right)\sin\left[2\frac{\Lambda}{H}\left(\frac{\eta-\eta_\text{in}}{\eta_\text{in}}\right)\right]\right) 
\nonumber\\
&+&\left[3\left(\frac{\eta}{\eta_\text{in}}\right)+2\left(\frac{\eta_\text{in}}{\eta}\right)-4\right]
\left[\text{Ei}\left(-2i\frac{\Lambda}{H}\left(\frac{\eta-\eta_\text{in}}{\eta_\text{in}}\right)\right)+\text{Ei}\left(2i\frac{\Lambda}{H}\left(\frac{\eta-\eta_\text{in}}{\eta_\text{in}}\right)\right)\right]
\bigg\},
\nonumber\\
&&
\end{eqnarray}
which, for $\Lambda\gg H$ and upon summing all the other possible combinations, gives
\begin{equation}\label{eq:finaldS}
\phi^2(\eta) = \frac{g^2}{8\pi^2}\left\{-\ln\left(\frac{\Lambda}{H}\right)
+\frac{1}{2}\frac{\sin\left[2\frac{\Lambda}{H}\left(\frac{\eta-\eta_\text{in}}{\eta_\text{in}}\right)\right]}{\frac{\Lambda}{H}
\left(\frac{\eta-\eta_\text{in}}{\eta}\right)
}
\right\}\chi^2(\eta).
\end{equation}
Notice that the natural renormalization scale $H$ appears now in the ultraviolet time-independent logarithm.  
The oscillating term with a fast decreasing amplitude quickly becomes small during inflation. To give a more quantitative argument, we will count the number of e-folds necessary to get an amplitude smaller than unity. The condition is (remember that $\eta<0$)
\be
\frac{\Lambda}{H}\left(\frac{\eta_\text{in}-\eta}{\eta}\right)\gg 1, 
\ee
that is 
\be
\frac{\eta_{\rm in }}{\eta}\ll \frac{H}{\Lambda}+1.
\ee
This quantity can be related to the number of e-folds $N$ 
\be
N = \ln(\eta_{\rm in}/\eta) = \ln\left(\frac{H}{\Lambda}+1\right) \sim \frac{H}{\Lambda}.
\ee
During inflation, because of the rapid expansion, the non-local term dies off after a fraction of an e-fold, and one is left with the mixing
$\phi^2(\eta) = -(g^2/8\pi^2)\ln(\Lambda/H)\chi^2(\eta)$. This mixing might be important if, for instance, we identify the field $\phi$ with the inflaton field driving inflation and the field $\chi$ with an extra light  field (this would require  the vacuum expectation value of the inflaton to be  small enough), whose energy density play a negligible role in the inflationary dynamics.
Nevertheless, if the field $\chi$ is highly non-Gaussian, then the non-Gaussianity  can be transferred to the inflaton field through the composite operator mixing and the three-point correlator might receive corrections $\sim (g^2/8\pi^2)^3\ln^3(\Lambda/H)$.

\section{Conclusions}
\noindent
In this paper we have investigated the phenomenon of composite operator renormalization and mixing in systems where
time-translational invariance is broken and the evolution is out-of-equilibrium. Through a simple toy model we have 
shown that composite operators mix also through  non-local memory terms which persist for periods whose duration is set by the 
mass scales in the problem. In particular, in the presence of large hierarchy among masses, the time length of the memory effects 
is typically dictated by the lighter mass. This renders the diagonalization of composite operators more difficult than it is in a time-independent 
setting. Our results my have interesting applications to many phenomena, such as baryogenesis and inflation, which took place in the
early universe. As observables are  expectation values of composite operators which  suffer operator mixing, one would expect that the memory kernels  play a role in the dynamics. 
We leave the study of these effects for future work.
 \section*{Acknowledgments}
We thank Richard Holman for useful comments on the draft and Peter Wittwer for help on some technicalities. A.R. is  supported by the Swiss National
Science Foundation (SNSF), project ``The non-Gaussian Universe" (project No. 200021140236).



\end{document}